# The Characterization of new $Eu^{2+}$ doped $TlSr_2I_5$ Scintillator Crystals

H. J. Kim[a*], Gul Rooh[b], Arshad Khan[a], H. Park[a], Sunghwan Kim[c]

[a]Department of Physics, Kyungpook National University, Daegu 41566, Korea
[b]Department of Physics, Abdul Wali Khan University, Mardan,23200, Pakistan
[c]Department of Radiological Science, Cheongju University, Cheongju 41566, Korea

[*] Corresponding author.
Tel.:+82-53-950-5323; fax: +82-53-956-1739. E-mail address: hongjoo@knu.ac.kr (H.J. Kim).

**Abstract**

$TlSr_2I_5$: $Eu^{2+}$ is a newly discovered scintillator and show promising scintillation properties for X- and γ-rays spectroscopy applications. Two zones vertical Bridgman technique is used for the growth of this scintillator. Luminescence properties of the grown crystals are measured under X-ray excitation at room temperature. Pure and $Eu^{2+}$ doped crystals contained broad emission bands between 445-670 nm peaking at 528 nm and 430-600 nm peaking at 463 nm, respectively. Energy resolution, light yield and decay time profiles are studied under 662 keV γ-ray excitation using $^{137}Cs$ radioactive source. Energy resolution of 4.2 % (FWHM) is obtained for 3 mol% $Eu^{2+}$ doped crystal. For the same sample, light yield of 70,000 ph/MeV is also obtained. Three and two decay constants are observed for the pure and $Eu^{2+}$ doped samples, respectively at room temperature. Effective Z-number and density are found to be 61 and 5.30 g/cm$^3$, respectively.

**Introduction**

Growing demand of excellent scintillators in different applications sparks the interest of researchers to discover new compounds. Various crystal groups are investigated for the search of new superior scintillators which includes binary, ternary and quaternary halide compounds. Among these, $SrI_2$: $Eu^{2+}$ [1], $LaBr_3$: $Ce^{3+}$ [2], $LaCl_3$: $Ce^{3+}$ [3], $CsBa_2I_5$: $Eu^{2+}$ [4, 5], $KSr_2I_5$: $Eu^{2+}$ [6], $Tl_2LaCl_5$: $Ce^{3+}$ [7], $Tl_2LaBr_5$: $Ce^{3+}$ [8], $TlMgCl_3$ [9], $TlCdCl_3$ [10], $Tl_2LiYCl_6$: $Ce^{3+}$ [11-13] are high performance scintillators, most of these scintillators are commercially available and are using in different applications. Although these scintillators exhibits high light yield, excellent energy resolution, fast scintillation response, moderate density and effective Z-number, however further enhancement of the scintillation performance could be possible by incorporating dense and high Z-number ions in these scintillators. For this purpose different strategies are employed and the replacement of cation by high Z-number ion is one of these which leads to the alteration of bandgap and positioning of the rare earth energy level in the hosts. For example, in $Cs_2LiYCl_6$: $Ce^{3+}$ crystal after replacing cesium (Cs) ion by relative high density and high Z-number thallium (Tl) ion, we found significant improvement in the scintillation properties of the $Tl_2LiYCl_6$: $Ce^{3+}$ [11]. Similarly, $Tl_2LiGdCl_6$: $Ce^{3+}$ [14] and $Tl_2LiLuCl_6$: $Ce^{3+}$ [15] showed promising improvements in scintillation performance after the replacement of Cs by Tl ion in $Cs_2LiGdCl_6$: $Ce^{3+}$ [16] and $Cs_2LiLuCl_6$: $Ce^{3+}$ [17] scintillators, respectively.

Ternary alkali alkaline-earth halide compounds with generic formula $XY_2Z_5$ (X = Li-Cs, Y = Ca-Ba and Z = Cl-I) are new potential scintillators and among these most are synthesized and reported earlier [4-6, 18, 19]. Recently, $KSr_2I_5$: $Eu^{2+}$ and $KBa_2I_5$: $Eu^{2+}$ are reported by Stand et al. [6, 20], these scintillators shows best scintillation results, however, the radioactive $^{40}K$ isotope in the host hinder the performance of these scintillators in different applications. It is worth mentioning here that the replacement of potassium (K) by Tl ion in $KSr_2I_5$ and $KBa_2I_5$ hosts not only resolve the radioactivity issue but also improve the effective Z-number ($Z_{eff\,(KSr2I5)}$ = 50, $Z_{eff\,(KBa2I5)}$ = 53, $Z_{eff\,(TlSr2I5)}$ = 61, $Z_{eff\,(TlBa2I5)}$ = 62) and density ($\rho_{KSr2I5}$ = 4.39 g/cm$^3$, $\rho_{KBa2I5}$ = 4.52 g/cm$^3$, $\rho_{TlSr2I5}$ = 5.30 g/cm$^3$) of these scintillators. Improvement in the effective Z-number and density of these scintillators ensuring higher detection efficiency of X- and γ-rays in many applications. Since we are working on different compounds of $TlY_2Z_5$ (Y = Ca-Ba and Z = Cl-I) and successfully discovered new efficient scintillators of this family which will be reported in our future work.

In this paper we present the results of new pure and $Eu^{2+}$ doped $TlSr_2I_5$ crystals for the radiation detection. This material is hygroscopic and grown from the melt using two zones vertical Bridgman method. Luminescence and scintillation properties measured at room temperature. Luminescence measurement include X-ray excitation while scintillation properties such as energy resolution, light yield and decay profile are measured under γ-ray excitation.

**Experimental Methods**

We used two zones vertical Bridgman technique for the growth of $TlSr_2I_5$: $Eu^{2+}$ single crystals. High purity (5N) anhydrous beads of TlI (Alpha-Aesar), $SrI_2$ (APL Engineered Materials) and $EuI_2$

(APL Engineered Materials) were used for the growth. Beads were mixed together according to the stoichiometric ratio and loaded into the quartz ampoules having inner diameter of 10 mm. Prior to sealing, all loaded ampoules were evacuated to $10^{-7}$ Torr and heated to ~200 °C for long time in order to remove the moisture contents from the mixed beads. The ampoules were then sealed and shifted to the Bridgman growth chamber. A vertical temperature gradient of 10 °C/cm between the two zones was used during the growth process. During growth, lowering speed of the ampoules was selected as 0.5 mm/h. After crystal growth, the grown crystals were cooled down to room temperature at the rate of 10 °C/h. The grown samples were transparent, free of inclusions as shown in the Fig. 1. Different dimension samples (~ 8 x 5 x 2 $mm^3$) were cut from the clean portion of the ampoules and polished for different characterizations. Due to the hygroscopic nature of this crystal, cutting, polishing and different characterizations were performed in ultra-dry environment. This scintillator has monoclinic crystal structure with lattice parameters a = 9.97 Å, b = 8.94 Å, and c = 14.25 Å [21]. The space group of this scintillator is $P2_1/C$ and density 5.30 $g/cm^3$. Effective Z-number was found to be 61.

Emission spectra measurements of $TlSr_2I_5$: $Eu^{2+}$ single crystals were performed under X-ray excitation source at room temperature. For the X-ray excited luminescence, an X-ray tube (DRGEM. Co.) having tungsten anode and operated at 50 kV and 1 mA power setting was used. The emission spectra of the irradiated crystals were measured using a spectrometer (QE65000 fiber optic spectrometer) made by Ocean Optics.

The gamma ray pulse height spectra were measured with a Hamamatsu R6233-100 photomultiplier tube (PMT) connected to a Canberra 2005 preamplifier and an Tennelec TC 245 spectroscopy amplifier. Crystals covered with several layers of Teflon tape were optically attached with the PMT window using optical grease and excited with 662 keV γ-ray from a $^{137}Cs$ source. Signals generated in the crystals were shaped with different shaping time and digitized using a 25-MHz flash analog-to-digital converter (FADC). The FADC output was recorded into a personal computer by using a USB2 connection and the recorded data was analyzed with a C++ data analysis program [22].

Scintillation decay profiles were measurement by optically coupled crystal with the PMT (Hamamatsu R6233-100) and excited with 662 keV γ-rays from a $^{137}Cs$ source. Output signals of the PMT were digitized using a 400-MHz FADC [23], which was fabricated to sample the pulse every 2.5 ns for duration up to 64 μs in order to fully reconstruct each photoelectron pulse [24].

The decay time profiles of the TlSr$_2$I$_5$: Eu$^{2+}$ single crystals were evaluated from the recorded pulse shape information.

## Results and analysis

### Luminescence spectra

X-ray excited luminescence spectra of pure and 3% Eu$^{2+}$ doped TlSr$_2$I$_5$ single crystals are shown in Fig. 2. Pure crystal shows broad emission band located between 445-670 nm. After deconvolution of the pure emission spectrum, we found three resolved bands with peaks located at 487 nm, 526 nm and 569 nm, respectively, see inset of Fig. 2. A recent theoretical study of Kang et al. [25] reported the formation of Tl-bound self-trapped exciton (STE) and the emission of Tl$^+$ ion around 477 nm in TlSr$_2$I$_5$ crystal. Moreover, in CsBa$_2$I$_5$: Tl$^+$ crystal, Tl$^+$ emission at 500 nm is also reported [26]. Therefore, the emission peak observed at 487 nm in the pure TlSr$_2$I$_5$ crystal is attributed to the $^3P_1 \rightarrow {}^1S_o$ transition of Tl$^+$ ions. Emission spectrum of the 3% Eu$^{2+}$ doped TlSr$_2$I$_5$ crystal shows broad band between 433-600 nm with the peak shifted in the lower wavelength region (blue shift) and found at 463 nm. This peak is assigned to the *$4f^65d^1 \rightarrow 4f^7$* transition of Eu$^{2+}$ ion. The broadness of the 3% Eu$^{2+}$ emission band appeared to be a superposition of several peaks and assigned to Tl$^+$ and STE emissions. Similar emission features are also reported in CsBa$_2$I$_5$: Eu$^{2+}$ [4] and are assigned to Eu$^{2+}$ and STE luminescence. Further investigations with varying Eu$^{2+}$ concentrations in TlSr$_2$I$_5$ crystal are underway for the concentration optimization and energy transfer mechanism from the host to Eu$^{2+}$ ion site.

### Pulse height spectra and light yield

Figure 3 shows the pulse height spectra of the pure and 3% Eu$^{2+}$ doped TlSr$_2$I$_5$ crystals. Crystals are excited with 662 keV γ-rays from a $^{137}$Cs source and recorded the pulse height spectra. The photopeak at 662 keV is fitted to a Gaussian function and calculated the energy resolution. The energy resolution of the 662 keV photopeak are 4.2% and 8.5 % full width at half maximum (FWHM) obtained for 3% Eu$^{2+}$ and pure crystals, respectively. Peaks located at low channels in Fig. 3 are due to the Tl K-X-ray escape peak [27].

Light yield measurement is performed by comparing the pulse height spectra of the TlSr$_2$I$_5$ crystals with that of a reference LYSO: Ce$^{3+}$ crystal (~33,000 ph/MeV) [28] under 662 keV γ-rays excitation from a $^{137}$Cs source by considering the quantum efficiency correction. Similar experimental setup of pulse height measurement as described in Experimental Methods is used for the light yield determination. Using identical conditions of PMT (Hamamatsu R6233-100) gain, shaping time and amplifier gain, pulse height spectra of all crystals are recorded under γ-rays excitation. After comparison of the photopeak channel numbers in the pulse height spectra, the highest light yield of 70,000 ph/MeV is observed for 3% Eu$^{2+}$ and 31,000 ph/MeV for pure crystal. Inset of Fig. 3 shows the pulse height spectrum of LYSO: Ce$^{3+}$ crystal. Comparing with NaI: Tl$^+$, BGO and LYSO crystals, TlSr$_2$I$_5$: Eu$^{2+}$ shows better scintillation properties and will perform better in different applications.

**Decay profiles**

Scintillation decay profiles of the pure and 3% Eu$^{2+}$ doped TlSr$_2$I$_5$ crystals are shown in Fig. 4. Decay profile of the pure and 3% Eu$^{2+}$ doped samples are well fitted by three and two exponential equations, respectively. The exponential fits, decay constants and their corresponding light yield contributions are shown in Fig. 3. More than one decay constants revealed different energy transfer mechanism from the host to the luminescence center in TlSr$_2$I$_5$ crystals. When a scintillator is excited by high ionizing radiation, electrons (e) and holes (h) are produced in the conduction and valance bands, respectively. These electrons and holes after thermalization follows different pathways through host lattice and deposit their energies to a luminescence center.

Previously, two energy transfer mechanisms of the e-h pairs from the host to Tl$^+$ ions were suggested in different iodide scintillators [29] which includes (I) hole capture and (II) electron capture, as given below.

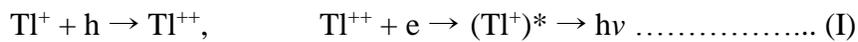
Tl$^+$ + h → Tl$^{++}$,      Tl$^{++}$ + e → (Tl$^+$)* → h$\nu$ ……………….. (I)

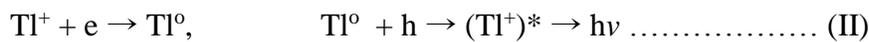
Tl$^+$ + e → Tl$^o$,      Tl$^o$ + h → (Tl$^+$)* → h$\nu$ ………………. (II)

In process-I, a hole is trapped by Tl$^+$ ion and produce Tl$^{++}$, while an electron is trapped by another Tl$^+$ ion to produce Tl$^o$. Subsequently, Tl$^o$ released an electron by thermal excitation and is trapped by Tl$^{++}$ ion to produce excited Tl$^+$ ion (Tl$^+$)* followed by the emission of photons due to the de-excitation of Tl$^+$ to ground state. Life time of Tl$^o$ at room temperature is long and therefore thermal release of electron is a slow process. A hole trapping is occurred through V$_k$ ($2I_2^-$ + h → V$_k$) center. A V$_k$ center can also trap an electron from the conduction band and form self-trapped excitons (STE) [25]. In process-II, electron is trapped at Tl$^+$ ion and produce Tl$^o$, a thermally excited hole is then trapped by Tl$^o$ and produce (Tl$^+$)* followed by the emission of photons due to the de-excitation of Tl$^+$ ion. Beside these two processes, direct e-h pair capture at Tl$^+$ ion to produce (Tl$^+$)* is also possible however, this process is independent of thermal excitation of electron and hole [30]. The decay constants obtained in the pure and 3% Eu$^{2+}$ doped TlSr$_2$I$_5$ crystals could be tentatively explained on the basis of the scintillation mechanism discussed in process-I and II. A relatively fast decay constant of 151 ns observed in the pure TlSr$_2$I$_5$ crystal (see Fig. 4) is ascribed to the Tl$^+$ ion emission due to the direct e-h capture at Tl$^+$ ion. Our recent study on Tl$_2$LaBr$_5$: Ce$^{3+}$[8] shows a similar decay constant observed in the pure sample due to Tl$^+$ ion. A decay constant in few hundred nano-seconds revealed that direct recombination of e-h pairs is not possible in 3% Eu$^{2+}$ doped TlSr$_2$I$_5$ crystal. Therefore, it is suggested that process-I and Process-II are responsible for the decay constants of 605 ns and 525 ns observed in the pure and 3% Eu$^{2+}$ samples, respectively. Since process-I and process-II are thermally activated processes, therefore, they significantly delayed the transfer of excitation energy (e-h pairs) from ionization track to Tl$^+$ sites; hence cause of the long decay constants observed in few hundred nano-seconds (605 ns and 525 ns). Long decay constants of 3-3.3 µs are attributed to the STE luminescence or some crystal defect or self-absorption affect [4, 15, 31, 32].

**Conclusions**

We reported the luminescence and scintillation properties of the newly developed TlSr$_2$I$_5$ crystals in pure and 3% Eu$^{2+}$ activated forms. This scintillator has a large effective Z-number and density. X-ray induced luminescence spectra of the pure and Eu$^{2+}$ doped samples exhibits Tl$^+$, Eu$^{2+}$ and STE emission bands between 445-670 nm and 433-600 nm, respectively. Best energy resolution of 4.2 % (FWHM) with 70,000 ph/MeV light yield are found for 3% Eu$^{2+}$ doped sample under 662 keV γ-ray excitation. Decay profiles of the investigated samples exhibits three and two exponential decay constants under γ-ray excitation at room temperature. A tentative scintillation mechanism described in ref. [25, 30] is proposed for the explanation of the observed decay constants of this scintillator. Compared with CsI:Tl$^+$ and NaI:Tl$^+$ crystals, the initial scintillation properties of TlSr$_2$I$_5$: Eu$^{2+}$ crystals are significantly higher and could be a potential new scintillator for X- and γ-rays detection in different applications. Better energy resolution (> 3%) at the observed light yield (70,000 ph/MeV) of this scintillator is expected however, the obtained energy resolution might be due to the impurity or crystal defects. Since, we are working on the purification of the material, doping of different mole% of $Eu^{2+}$ ion and also optimization of the crystal growth conditions of this scintillator, therefore, further improvement in the scintillation performance is expected in future.

**Acknowledgment**

These investigations have been supported by the National Research Foundation of Korea (NRF) funded by the Ministry of Science and Technology, Korea (MEST) (No. 2015R1A2A1A13001843).

**References**

[1]     Y. T. Wu, L. A. Boatner, A. C. Lindsey, M. Zhuravleva, S. Jones, J. D. II. Auxier, H. L. Hall, C. L. Melcher, Cryst. Growth Des., 15 (2015) 3929.

[2]     M.S. Alekhin, J.T.M. de Haas, I.V. Khodyuk, K.W. Krämer, P.R. Menge, V. Ouspenski, P. Dorenbos, Appl. Phys. Lett. 102 (2013) 161915.

[3]     H. B. Chen, P. Z. Yang, C. Y. Zhou, C. Y. Jiang, J. G. Pan, Cryst. Growth Des., 6 (2006) 809.

[4]     M. S. Alekhin, D. A. Biner, K. W.Kramer, P. Dorenbos, J. Lumin. 145 (2014) 723 .

[5]     E. D. Bourret-Courchesne, G. Bizarri, R. Borade, Z. Yan, S. M. Hanrahan, G. Gundiah, A. Chaudhry, A. Canning, SE. Derenzo, Nucl. Instrum. Methods Phys. Res. A 612 (2009) 138.

[6]     L. Stand, M. Zhuravleva, A. Lindsey, C. L. Melcher, Nucl. Instrum. Methods Phys. Res. A 780 (2015) 40.

[7]     H. J. Kim, Gul Rooh, Sunghwan Kim, J. Lumin. 186 (2017) 219.

[8]     H. J. Kim, Gul Rooh, Arshad Khan, Sunghwan Kim, Nucl. Instrum. Methods Phys. Res. A 849 (2017) 72.

[9]     Y. Fujimoto, M. Koshimizu, T. Yanagida, G. Okada, K. Saeki, K. Asai, Jpn. J. Appl. Phys. 55 (090301) (2016) 090301.

[10]    Y. Fujimoto, K. Saeki, T. Yanagida, M. Koshimizua, K. Asai, Radiation Measurements (2017), http://dx.doi.org/10.1016/j.radmeas.2017.03.034 (in press).

[11]    H. J. Kim, G. Rooh, H. Park, S. Kim, IEEE Trans. Nucl. Sci., 63(2) (2016) 439.

[12] R. Hawrami, E. Ariesanti, L. Soundara-Pandian, J. Glodo, K. S. Shah, IEEE Trans. Nucl. Sci., 63(6) (2016) 2838.

[13] G. Rooh, H. J. Kim, H. Park, S. Kim, J. Cryst. Growth 459 (2017) 163.

[14] H. J. Kim, G. Rooh, H. Park, S. Kim, J. Lumin. 164 (2015) 86.

[15] G. Rooh, H. J. Kim, Jonghun Jang, Sunghwan Kim, J. Lumin. 187 (2017) 347.

[16] G. Rooh, H. J. Kim, H. Park, Sunghwan Kim, J. Cryst. Growth 312 (2010) 2243.

[17] A. Bessiere, P. Dorenbos, C.W.E. van Eijk, K.W. Kramer, H.U. Gudel, A. Galtay- ries, J. Lumin. 117 (2006) 187.

[18] H. P. Beck, G. Clicque, H. Nau, Z. Anorg. Allg. Chem. 536 (1986) 35.

[19] C. M. Fang , Koushik Biswas, J. Phys. Chem. C  120 (2016) 1225.

[20] L. Stand, M. Zhuravleva, B. Chakoumakos, J. Johnson, A. Lindsey, C. L. Melcher, J. Lumin. 169 (2016) 301.

[21] G. Schilling, G. Meyer, *Z. Anorg. Allg. Chem.*, 622 (1996) 759-765.

[22] J. H. So, H. J. Kim, H. Kang, H. Park, and S. Lee, J. Korean Phys. Soc. 52(3) (2008) 925.

[23] Online: http://materials.springer.com/isp/crystallographic/docs/sd_1711773.

[24] H. J. Kim et al. IEEE Trans. Nucl. Sci., 57 (3) (2010) 1475.

[25] B. Kang, C. M.  Fang, K. Biswas, J. Phys. D: Appl. Phys. 49 (2016) 395103.

[26] M. Gascon, E.C.Samulon, G. Gundiah, Z. Yan, I. V. Khodyuk, S.E. Derenzo, G. A. Bizarri, E.D.J. Bourret-Courchesne,  J. Lumin. 156 (2014) 632014.

[27] H.J. Kim, G. Rooh, H. Park, S. Kim, J. Lumin. 164, 86 (2015).

[28] G. Rooh, H. J. Kim, H. Park, S. Kim, J. Cryst. Growth 312 (2010) 2243.

[29] P. A. Rodnyi, Physical Processes in Inorganic Scintillators. CRC Press; Boca Raton, New York: 1997.

[30] R. B. Murray, IEEE Trans. Nucl. Sci., NS-22 (1975) 54.

[31] M.D. Birowosuto, P. Dorenbos, Phys. Status Solidi A 206 (2009) 9.

[32] Y. Sakai, M. Kawahigashi, T. Minami, T. Inoue, S. Hirayama, J. Lumin 42 (1989) 317.

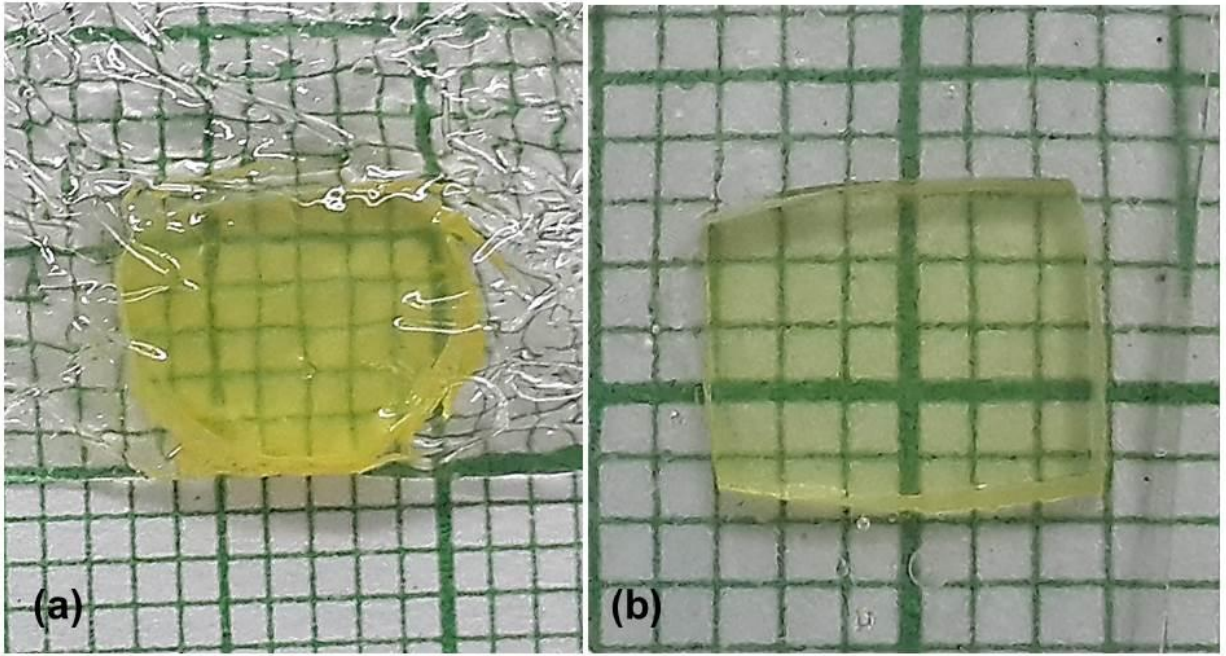

Figure 1. Photograph of the cut and polished (a) Pure and (b) 3 mol% Eu$^{2+}$ doped TlSr$_2$I$_5$ crystals.

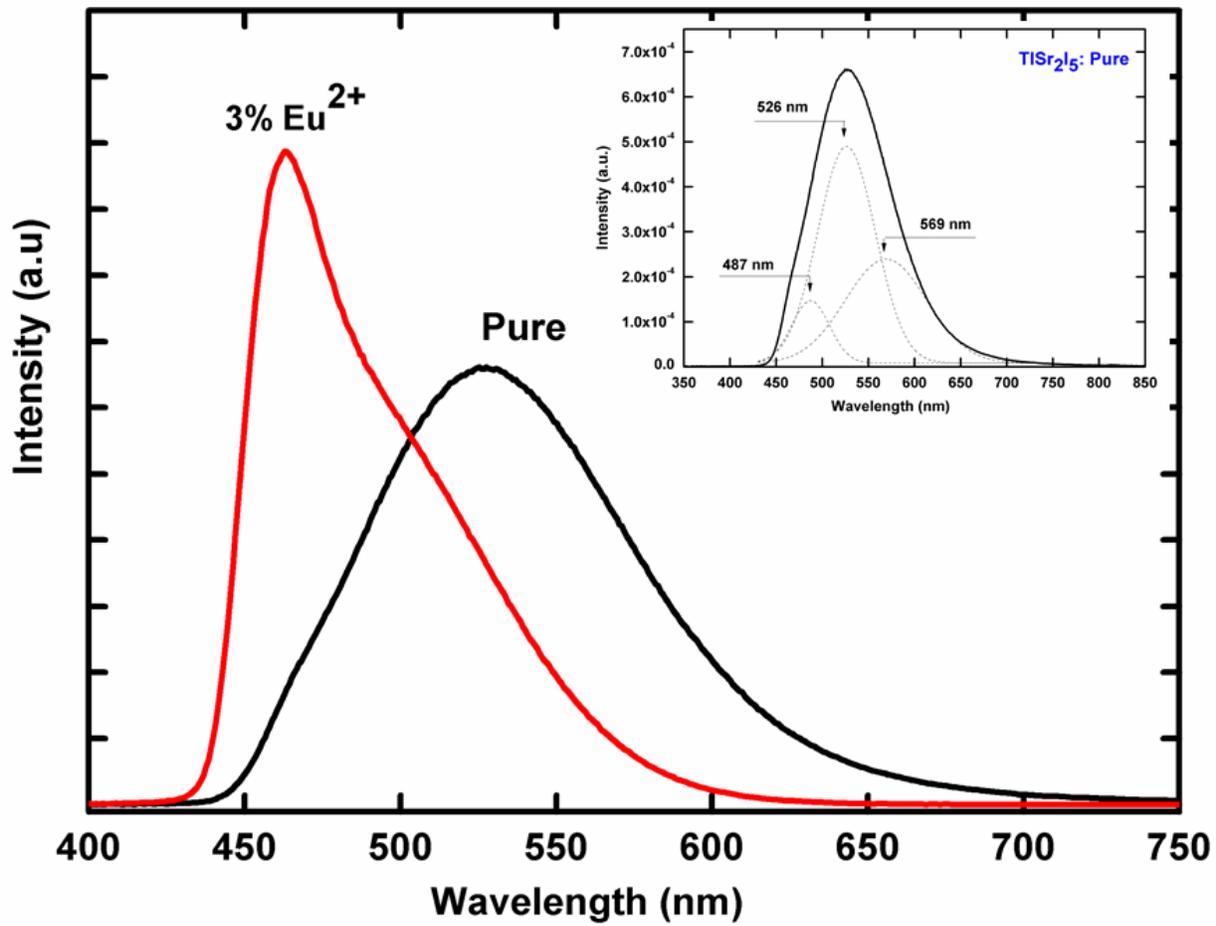

Figure 2. X-ray induced emission spectra of pure and 3 mol% $Eu^{2+}$ doped crystals of $TlSr_2I_5$ at room temperature.

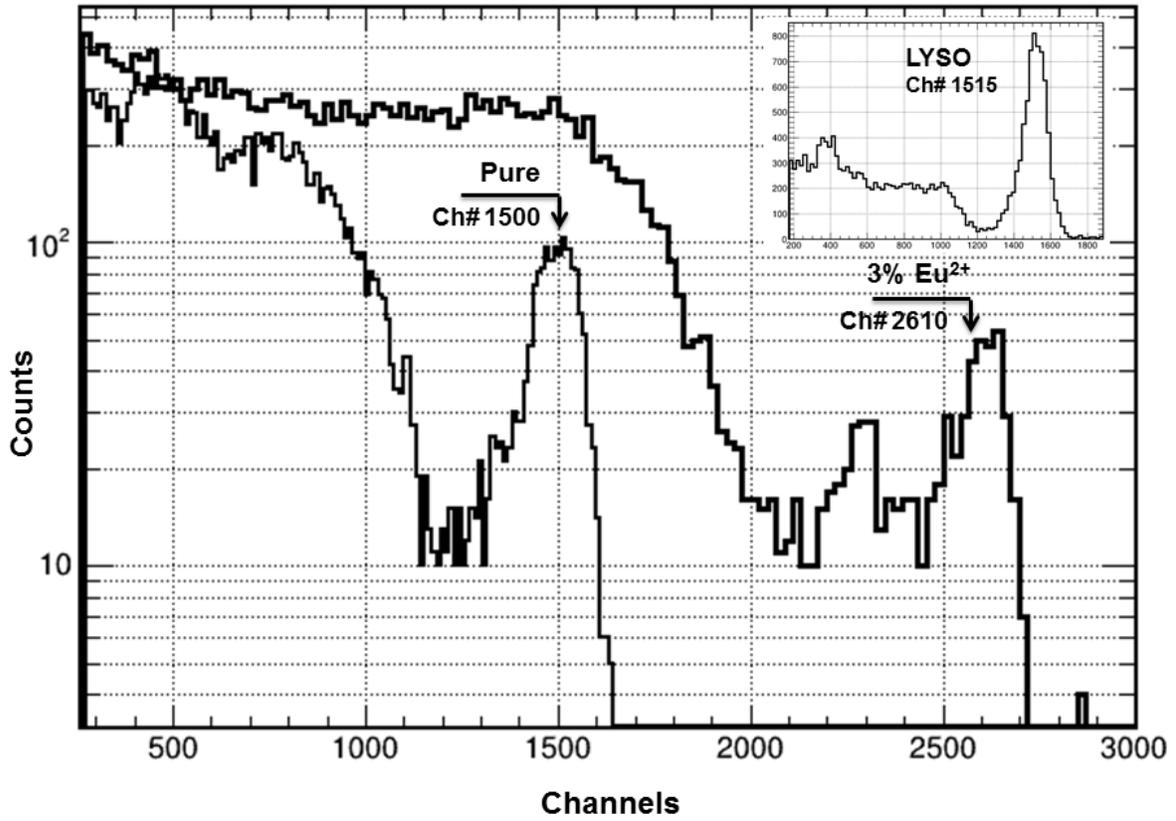

Figure 3. Scintillation pulse height spectra of pure and 3% $Eu^{2+}$ doped $TlSr_2I_5$ single crystals irradiated with γ-rays from a $^{137}Cs$ source. Inset shows the pulse height spectrum of LYSO: $Ce^{3+}$ crystal.

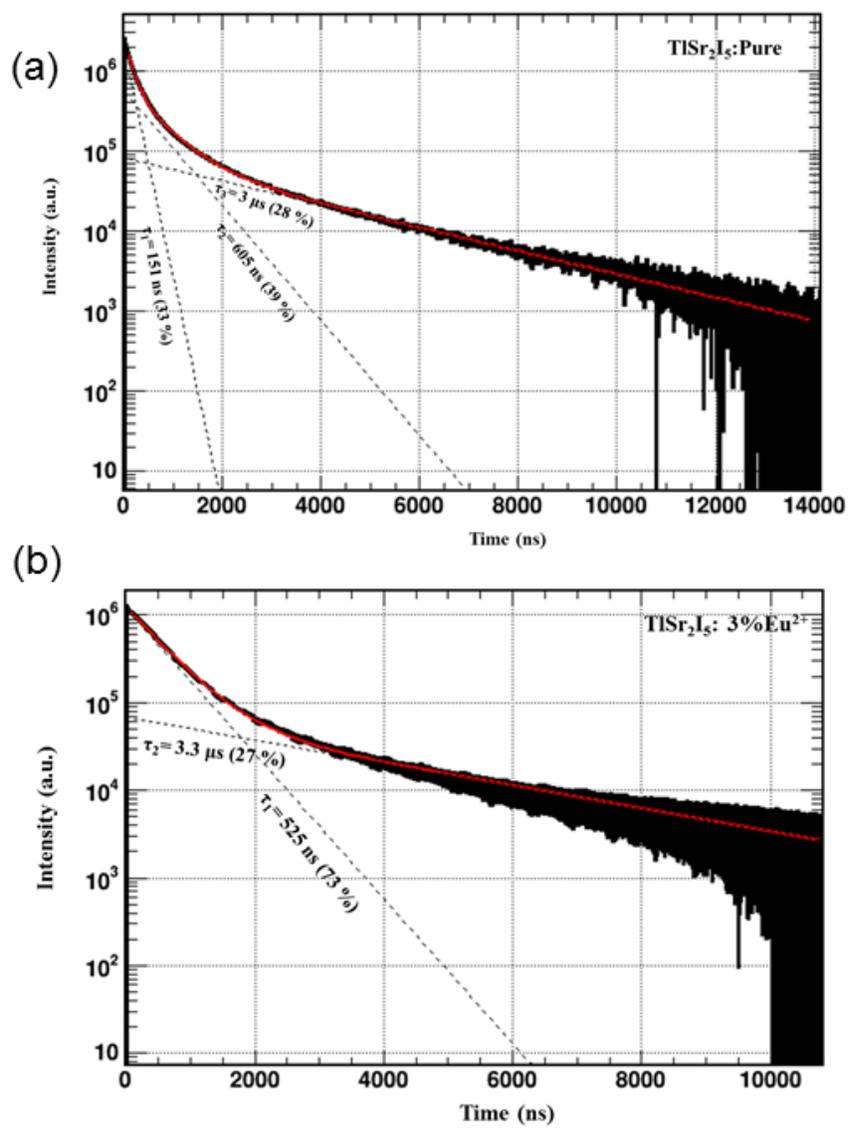

Figure 4. Scintillation decay profiles of a) pure and b) 3% $Eu^{2+}$ doped $TlSr_2I_5$ crystals measured under γ-rays excitation from a $^{137}Cs$ source at room temperature.